# Application Behavior Enforcement Based On Network Characteristics

Pinaki Mitra
Indian Institute Of Technology,
Guwahati pinaki@iitg.ernet.in

Girish Sundaram
IBM Information Management,
India Software Lab
gisundar@in.ibm.com

Swarup Kumar Mallick
Indian Institute Of Technology,
Guwahati
swarup@iitg.ernet.in

*Abstract*—Every device defines the behaviour of various applications running on it with the help of user profiles or settings. What if a user wants different different applications to run in different different networks. Then he can't do this. Because in today's existing operating systems the application profile, user specific tool settings and any such system usage settings do not consider the current network characteristics of the user/system and are therefore static in nature. Therefore there is a need for an intelligent system which will dynamically change or apply changes to the settings of various applications running on the system considering the current network characteristics based on user specifications. This paper presents an idea such that a user can set different different applications in different different networks. This paper presents how the user will get pop up messages when he visits to safe web sites. And according to the user's current network status he will get a pop up message when he will go for download or stream audio or video files.

## I. INTRODUCTION

Every device defines the behaviour of various applications running on it with the help of user profiles or settings. For example A user can set the default web page when the web browser opens, he can specify that whenever he logs on to the machine the chat services should automatically start with him in the Available mode etc. But the OS can't automatically load the user specific application settings based on the network characteristics.

For example an employee (Joe) carrying a laptop may be in an Office Network in the day time from 9:00 am to 6:00 pm and will be at Home Network after that. He will continue to use the same laptop at both these locations. Since the user profiles and application settings are static in nature they will continue to behave irrespective of the network settings of the individual.

To elaborate further, Joe would have http://www.office.com as a default web page and would have enabled automatic log on to Lotus Sametime Server for chat server. These settings are perfect as long as he is in Office Network but when he comes back home he would need to manually modify the default settings of these applications or would need to manually start the needed applications (like he may want to log on to google talk and open gmail.com for checking his personal emails). That means this OS can't automatically load the user specific application settings based on the network characteristics.

Therefore there is a need for an intelligent system which will dynamically apply changes to the settings of various applications running on the system considering the current network characteristics based on user specifications.

## II. CORE IDEA

The Idea talks about a system which would be able to identify the characteristics of the network that the user device(laptop, palmtop etc) has connected to. This we can find out using any existing mechanism(e.g. ifconfig), which will help in differentiating the various networks he connects to. Check the current network exists in the Network Repository or not(Network Repository stores the mapping between various networks and their corresponding application settings which are given by user). If a match for the current network exists in Network Repository then go ahead and apply the user settings for various selected applications. Whenever the user connects to a new network (which means the user connects to a network which is not in the Network Repository) the system will prompt the user for information and the user can make an entry in the Network Repository and also at the same time also choose various applications for which the user wants to apply some specific settings.

For example while in Office Network the user wants http://www.office.com as the default web page, Lotus Sametime as default Instant Messaging Software etc. While at Home Network he wants http://www.google.com as the default web page, Google talk as default Instant Messaging Software etc. To handle this the Network Repository will store the user settings in a file named the network address(like DNS server information, IP Address etc on which a differentiation between various networks can be made). When the user connects to a network then the system will find the network details from the Network Repository if the current network exists then the system will find the user settings from that file and choose various applications for which the user wants to apply some specific settings.

### A. How can this idea be implemented?

Following are the various components of this invention.

- Network Daemon : This concept is used to get the current network of the system continually. According to the current network the Modifier Module will be called.

- Network Repository : This is a repository which has the data for various networks and their corresponding settings as specified by the user and this can exist on the same system that the user is using or can be any other system which is accessible to the user's system.
- Modifier Module : This module applies the user settings for a specific network to various chosen applications.
- User Interface : This is for the user to submit his choices for different applications. This is called when the user connects to a network which is not present in the Network Repository.

B. Algorithm to implement this Idea:

1: while true do
2:   if Network Daemon receives an address which is not present in Network Repository then
3:     The user will get a User Interface to give his choices for the new network
4:     Store those informations in the Network Repository and in the Cache
5:     Apply the settings for the current location from the Cache
6:   else
7:     if Network Daemon receives an address same as the current set address then
8:       Don't do anything
9:     else
10:       if The current address exists in the Cache then
11:         Retrieve the corresponding data from Cache and apply the settings
12:       else
13:         Go to the Network Repository in secondary memory, retrieve the data, store it in Cache and apply the settings
14:       end if
15:     end if
16:   end if
17: end while

## III. IMPLEMENTATION DETAILS

### A. Network Daemon

This concept is used to get the current network of the system at every moment. A program called daemon.c which is a daemon means this program runs in the background. When the System starts it will start automatically. The program use the existing command ifconfig. From this it will get the current ipaddress of the system. Then it will check whether the Link Encapsulation is Ethernet or Point-to-Point. If Link Encapsulation is Point-to-Point then the ipaddress will be the file name in Network Repository and all user specific settings will be stored in that file. If Link Encapsulation is Ethernet then using iwconfig it can be confirmed whether it is wired connection or wireless connection. In both these cases the ipaddress will concatenate with the DNS-server address to make the file name. The DNS-Server address will available from /etc/resolv.conf.

### B. Network Repository

This is the place where all the user specific settings corresponding to a network(i.e. IP Address) are stored. When the user first time installed this software in it's system then a hidden folder named .networkdaemon will be created on the Home Folder i.e. /home/username. In this folder all user specific settings will be stored after onwards. When the user connects to a network then the program will first search whether there is a file exists corresponding to the current network. If exists then it will read from that file, otherwise it will prompt a User Interface(UI) for the user input corresponding to that network and store the user inputs in a file named that new network.

### C. User Interface

This program is called when the user connects to a network which is not present in the Network Repository. This program takes a command line argument. This argument is the current IP Address, which will be shown in the UI. When the user will submit this form then this program will create a file named that IP Address and store all the data given by the user in that file and store this file in Network Repository.

### D. Browser Settings

For Browser Setting first search for the Profile folder of the browser e.g. for firefox it is .mozilla which is a hidden folder in Home Folder. Find the file which stores all the settings, for firefox it is prefs.js. How to get this file? First search for .mozilla folder in Home Folder. Here .mozilla is a hidden folder. Within that folder search for the folder firefox. In firefox there is a file named profiles.ini. In that file there is a setting for Path. Find the folder named that path name. In that folder search for a file prefs.js which is the desired file. In that file search for the phrase user_pref("browser.startup.homepage", any url), here just change the url part then save it. Now when the user connects to firefox then the corresponding url will open as the home page.

### E. Desktop Application Settings

This setting is useful for audio, video or any text based applications. Here the example is given to handle the video part that means which media player will be the default media player for a corresponding network. Changes can be made in two places.

- /usr/share/applications/defaults.list
- /home/username/.local/share/applications/mimeapps.list

In UBUNTU 11.04 version the local folder is searched first. We should update in the local folder. So first search for the mimeapps.list then search for the phrase video/X=P.desktop. Here X is the format of the video file e.g. mp4, dv, flv, 3gpp etc, and P is the media player name e.g. vlc, totem, real etc. We only have to change the P part for the default video player. This data we can get from the Network Repository. After changing the P part when we open a video file it will open with the same player that we have set in mimeapps.list file.

## F. Instant Messanger Setting :

A user has many accounts e.g. facebook,google talk i.e gmail, yahoo etc. Here the task is to activate one of these account in available mode. Which account to activate according to the network that data will available from the Network Repository. Here the software pidgin has used which is easy to use and free chat client software.

How to do this :

- First find the profile directory of the installed software. e.g. for pidgin the profile directory is

  /home/username/.purple . Here .purple is a hidden folder which will be automatically created when someone install pidgin.

- In .purple find a xml file accounts.xml .
- In accounts.xml search $<$ account $>$ wise. First go to 1st $<$ account $>$ field then search for $<$ name $>$ field. Then check whether the corresponding account matches with your data . If no match found then again search start from next $<$ account $>$ field . For example if you want to open facebook chat service then you should search for $<$ name $>$someaccountname@chat.facebook.com/ $<$ /name $>$.

- If match found then within that $<$ account $>$ and $<$ /account$>$ field search for $<$ statuses $>$ field. Then search for

  $<$ status type $=^0$ available$^0$ name $=^0$ Available$^0$ active $=^0$ false$^0 >$

  $<$ attributes/$>$

  $<$ /status$>$

  Here if active $=^0$ false$^0$ then make it $^0$true$^0$.
- Then find

  $<$ settings ui $=^0$ gtk $-$ gaim$^0 >$

  $<$ setting name $=^0$ auto $-$login$^0$ type $=^0$ bool$^0 > 0 <$ /setting$>$

  $<$ /settings$>$

  Here make the 0 to 1 then just save it.

## G. Email Program Setting

In this section call a command through the system call in the program. This data will available from the Network Repository. For example if user want to open Thunderbird then through system call just call thunderbird (Thunderbird should be installed in your system).

## H. Giving Pop Up Message When Visiting Safe Sites

The packet capture library libpcap has been used to achieve this task. Packet Capture, simply means to grab packets. Using this library we can grab all packets of a network. Here first we have to check all outgoing and incoming tcp packets whose data fields are not empty, that means which have some application data. To know a packet is tcp or not we have to check the IP headers protocol number. If this protocol number is 6 then the packet is a tcp packet. For checking outgoing packets we have to compare the packet's ethernet source address with the system's hardware address. If those two addresses match then the packet is an outgoing packet. Similarly for checking incoming packets we have to compare the packet's ethernet destination address with the system's hardware address. If those two addresses match then the packet is an incoming packet. For checking the data field is empty or not we have to subtract the total header length from the payload of the packet, if the result greater than 0(ZERO) then the packet has some application data. Here the total header length is the sum of Ethernet Header and IP Header. Now in the desired outgoing packet we have to check the HTTP Method or the Port Number. Here we will only consider the packets which have HTTP Method used is CONNECT or the outgoing Port Number is 443. We can check only the HTTP Method but in some cases the Port Number required. Similarly we can check only the Port Number but in some cases the HTTP Method required. Find the Acknowledgement Number of the outgoing packet which satisfy the above criteria, then retrieve the URL from the application data of that packet and save them correspondingly. Again in the desired incoming packet first find the sequence number of that packet. From the property of Transmission Control Protocol we know that if the Acknowledgment Number of a packet is equal to the Sequence Number of another packet then these two packets are Request and corresponding Response Packets. So here we will compare the Sequence Number of the incoming packet with all the Acknowledgment Numbers we have. If any one Acknowledgment Number matches then the current incoming packet is a Response Packet corresponding to the outgoing packet whose Acknowledgment Number matches with that sequence number. If no Acknowledgment Number matches with that Sequence Number then we don't need to go further test for this packet. Because this packet is a Response Packet but the request corresponding to this response packet has not used SSL Connection. So we don't need that packet. Now when the Acknowledgment Number and Sequence Number matches then check for response message in the incoming packet. First check the HTTP Response Code is 200 or not. If the HTTP Response Code is 200 then go further test for this packet otherwise don't need to go further test for this packet. Here in the response message we have to check whether there is a phrase like HTTP/1.0 200 Connection established or HTTP/1.1 200 Connection established. If there is a phrase like that in the response message then retrieve the URL corresponding to that Acknowledgment Number. Now this URL must be a https url and we can give a pop up message to the user that this site is a safe site. But there are certain issues regarding this. The first issue is when we connect to a site then it internally connects to other sites with https protocol. So in that case we will give pop up message for every url it connects. But we should not do that. We should only give the pop up message for the site which the user has entered. The second issue is when we connect to a site then that url is sent as request more than once. But we should not give the pop up message more than once.

*1) How To Solve These Issues:* For solving those issues we will take help of mozilla's places.sqlite file. Here we will use two tables of places.sqlite. Those are moz_places and moz_historyvisits. First we will take an integer array which will store the session of a page. Now when a url come then we will find it's most recent session. Then check whether it is already present in the array. If it is then we will not give a pop up message to the user. If it is not preseent then we will store it in the array and will give a pop up message to the user. In this way the two issues can be resolved. Because in the first case when we connect to a site then it internally connects to other sites with https protocol. But here all the other sites have the same session as the main site. So we will give pop up message for only one time. In the second case when we connect to a site that url is sent as request more than once. But every time the session will be same and it is already there in the array. So we will not give the pop up message more than once. In this case if we will open a new tab with the same url then it has a different session so now the pop up message will come.

*2) Algorithm To Implement This:*

1: **while** true **do**
2:   **if** Packet's Ethernet Source Address = System's Hardware Address **then**
3:     **if** Packet's Protocol Number = 6 **then**
4:       Goto step 12
5:     **end if**
6:   **end if**
7:   **if** Packet's Ethernet Destination Address = System's Hardware Address **then**
8:     **if** Packet's Protocol Number = 6 **then**
9:       Goto step 15
10:     **end if**
11:   **end if**
12:   **if** HTTP Method used = CONNECT OR Destination Port Number = 443 **then**
13:     Store the requested URL from the application data and the corresponding Acknowledgement Number of that Packet
14:   **end if**
15:   **if** Packet's Sequence Number matches with any one of the Acknowledgement Number we have stored **then**
16:     **if** HTTP Response Code = 200 Connection established **then**
17:       **if** The SessionId Of the URL corresponding to this Acknowledgement Number exists **then**
18:         Give a pop up message to the user that tis URL is a safe site
19:       **end if**
20:     **end if**
21:   **end if**
22: **end while**

## I. Giving Pop Up Message While Download Or Stream Audio Or Video Files Other Than Home Network

The packet capture library libpcap has been used to achieve this task. Packet Capture, simply means to grab packets. Using this library we can grab all packets of a network. Here we have to check all incoming tcp packets whose data fields are not empty, that means which have some application data. To know a packet is tcp or not we have to check the IP headers protocol number. If this protocol number is 6 then the packet is a tcp packet. For checking incoming packets we have to compare the packet's ethernet destination address with the system's hardware address. If those two addresses match then the packet is an incoming packet. For checking the data field is empty or not we have to subtract the total header length from the payload of the packet, if the result greater than 0(ZERO) then the packet has some application data. Here the total header length is the sum of Ethernet Header and IP Header. Here first we will check whether the network is Home network or anything else. If the network is Home network then we will not do anything. If the network is other than the Home network then we have to check the application data of every incoming packet. Here in application data we will search for the HTTP header Content-Type. This header indicates the Internet media type of the message content, consisting of a type and subtype. Here in the type or subtype field we have to check the MIME type for audio or video files. Here in the Content-type header we have to check the type and/or subtype field. If the type field is audio or video then this is a audio or video file. The other case can arise when the type field is application or x-music then we have to check the subtype field. If the phrase in the subtype field matches with the MIME type for audio or video files then this must be a audio or video file. So in those two cases we will give a pop up message to the user. There is an issue regarding to this problem. The issue is when we stream a file then how to know which packets are corresponding to that particular file. Otherwise every time when the Content-Type matches with audio or video type then we will give pop up message to the user which is undesirable. When we stream or download a audio or video file, all the contents come through one port. So we need to know the port number through which the connection has established. Then for every unique port we can give a pop up message to the user.

1) How To Solve This Issue: Here we are considering only the incoming packets. First check the HTTP Response Code is 200 OK or not. This condition is required because in some cases the Response code is 204 No Content but the MIME type matches with audio or video type, but we should not give the pop up message in those cases. So we should avoid those situations. If the response code is 200 OK check the MIME type matches with audio or video type. If matches then check whether there is a header named Content-Length exists or not. If the Content-Length header field exists then this is our desired packet. Now store the destination port number of this packet. Because all the packets corresponding to that streaming file will come through one port. So we will give pop up message only when we get a new port number.

2) Algorithm To Implement This:

```
1:  while true do
2:    if Packet's Ethernet Destination Address = System's
      Hardware Address then
3:      if Packet's Protocol Number = 6 then
4:        if HTTP Response Code = 200 OK then
5:          if Content-Type = audio or video type then
6:            if Content-Length field exists then
7:              Store the Packet's destination Port Number
8:              Corresponding to unique Port Number give
                a pop up message to the user
9:            end if
10:         end if
11:       end if
12:     end if
13:   end if
14: end while
```

### J. Caching Of Networks

When we use some particular networks in some regular intervals then there should be a caching mechanism. Because we should not every time go to the secondary memory and retrieve the settings for a particular network. For that reason we have implemented a cache which will store a limited number of networks and it's settings. This cache has implemented using LRU(Least Recently Used) method. For this cache we will not every time go to the secondary memory to get the application settings of a particular network. First we will search in the cache, if that particular network exists in cache then we can get the application settings from the cache. Otherwise we will go to the secondary memory to retrieve the data then store a copy of this data in the cache.

## IV. CONCLUSION AND FUTURE WORK

Using this idea the user will be no longer required to make any manual modifications to his settings he has stored for various applications in his system. The program will automatically find out the current network information of the system (user) using any of the existing mechanisms(like ifconfig, iwconfig or any other OS specific commands). Find out the user specified settings for that specific network and then make the needed modifications to the settings of the various applications running on it. A network database is a repository which has the data for various networks and their corresponding settings as specified by the user and this can exist on the same device that the user is using or can be any other system which is accessible to the user's system.

Here we are giving pop up message when the user download or stream audio or video files. It can be extended such that we can block the streaming or downloading. This we can do using iptables. Here we are opening the email program automatically. It can be extended such that when the user is in office network he should get only office related mails and when the user is in home network he should get all mails.